\documentclass[12pt]{article}

\begin{document}

\begin{center}
\textbf{O.A.Khrustalev, M.V.Tchitchikina}

\textbf{QUANTUM GRAVITY on the }

\textbf{CLASSICAL BACKGROUND: GROUP ANALYSIS, Part I }
\end{center}

\section{Introduction}

{\normalsize In our article we present an approach to the description of
gravitational field from the point of view of quantum field theory. We
consider quantization in the neighbourhood of solution of Einstein equation
and use the method of Bogoliubov transformations for this purpose. }

{\normalsize Quantum gravity is a pure theory and couldn't be tested by
laboratory experiments and astrophysical data, because in observable
processes of Universe quantum effects connected with gravity are extremely
small. In the same time gravitational interaction is universal one for all
kinds of matter independently from it properties. Quantum gravity is not
constructed yet, and in the present moment there are some relatively
independent directions: }

{\normalsize quantum theory of gravitational field; }

{\normalsize theory of nongravitational field in the curved space-time; }

{\normalsize quantum cosmology and quantum theory of black holes; }

{\normalsize quantum supergravity and multidimensional theory of unification
of interactions. }

{\normalsize Quantum theory of gravitational field is based on the
quantization of classical theory of gravitational field - general
relativity. }

{\normalsize Theory of the field in the curved space-time investigates the
methods of quantization of matter fields on the background of classical
gravitational field. }

{\normalsize Quantum cosmology is application of methods of quantum gravity
to the Early Universe in the vicinity of singularity. The most important
achievement of quantum cosmology is construction of concrete models if
inflationary universe. }

{\normalsize Quantum theory of Black holes investigates mainly the effects
of particles creations and vacuum polarization of the gravitational field of
black holes. }

{\normalsize Quantum gravity is close to the multidimensional theories of
Grand Unification. Unification of space-time symmetries with inner
symmetries and gauge symmetries of strong and electroweak interactions is
reached due to introduction of curves space-time of 4+d dimensions. The
symmetry of this space-time determines the symmetry of interactions. }

{\normalsize In future all of those directions are believed to be the parts
of Unified quantum gravity theory. That is why the appropriate choice of the
methods of quantization of gravitational field seems to be very important. }

{\normalsize Quantization on the classical background seems to be a very
special case, but existence of classical component is an essential feature
of gravitational field and our choice of quantization looks to be
reasonable. }

{\normalsize We consider exact solutions of Einstein equation to be
classical background and quantize in the neighbourhood of this exact
solutions. }

{\normalsize Any space-time could be treated as a solution of Einstein
equation 
\[
R_{ab}-{\frac 12}Rg_{ab}+\Lambda _{ab}=8\pi T_{ab} 
\]
with the appropriate choice of energy momentum tensor $T_{ab}$ of some
concrete form of the matter. }

{\normalsize However it is possible to find exact solutions only for the
space with a rather high degree of symmetry. Apart from this fact exact
solutions describes some kind of ideal situation: any origin of space-time
contains different types of matter. And it is possible to obtain exact
solutions only in the case of simple enclosed matter. }

{\normalsize Nevertheless exact solutions are important, because they give
ideas concerning qualificationly new phenomena that could arise in the
general relativity and hence point on the possible properties of real
solutions of field equations. The following examples shows a lot of
interesting types of solutions. }

{\normalsize a) The simplest metrix has constant curvature. It means that
the space is homogeneous. }

{\normalsize If $R>0$ we have De Sitter space-time of the first type, that
describes space-time of stationary Universe model. }

{\normalsize If we believe that our Universe is approximately homogeneous
and spherically symmetrical in large scale, than we have a model of
Robertson-Walker (or Freedman) and the metrix reads: 
\[
ds^2=-dt^2+S^2(t)d\sigma ^2, 
\]
here $d\sigma ^2$ is 3D metrix of space with constant curvature, the
geometry of space depends on sine of 3D curvature. }

{\normalsize Robertson-Walker solution symmetry demands energy -momentum
tensor to be the tensor of ideal fluid that describe matter in Universe
spread homogeneously. }

{\normalsize Large-scale solutions are good model for large scale matter
distribution, but for the single objects like stars we have to use
asymptotically flat solutions. }

{\normalsize b)This geometry with good approximations could be described by
well known Schwartzshild solution that represent spherically symmetrical out
space of massive body. }

{\normalsize 
\[
ds^2=-\left( 1-{\frac{2m}r}\right) dt^2+\left( 1-{\frac{2m}r}\right)
^{-1}dr^2+r^2d\Omega ^2 
\]
}

{\normalsize c)Out space-time of charged spherically symmetrical body is
described by Reisner-Nordstrem solution.(Though it has no spin neither
magnetic angular momentum hence it couldn't be good description of
gravitational field of electron.) This is unique asymptotically flat
solution of Einstein-Maxwell equation: }

{\normalsize 
\[
ds^2=-\left( 1-{\frac{2m}r}+{\frac{e^2}{r^2}}\right) dt^2+\left( 1-{\frac{2m}%
r}+{\frac{e^2}{r^2}}\right) ^{-1}dr^2+r^2d\Omega ^2 
\]
}

{\normalsize d)Kerr solution describe stationary axial-symmetrical
asymptotically flat field outside the rotating massive object: }

{\normalsize 
\[
ds^2=\rho ^2\left( {\frac{dr^2}\Delta }+d\Theta ^2\right) +\left(
r^2+a^2\right) \sin ^2\theta d\Phi -dt^2+{\frac{2mr}{\rho ^2}}\left( a\sin
^2\Theta d\Phi -dt\right) ^2 
\]
}

{\normalsize Kerr solution seems to be unequally possible solution for black
holes. }

{\normalsize e)Another interesting solution is a solution of flat waves in
the empty space-time and the metrix reads: }

{\normalsize 
\[
ds^2=2dudv+dy^2+dz^2+H(x,y,z)du^2 
\]
}

{\normalsize Those physically interesting exact solutions could be chosen as
a classical background. There are a lot of other exact solutions but not
much of them are studied perfectly. }

{\normalsize Now we would like to give some arguments why the method of
Bogoliubov group variables have advantages while we quantize the field with
some type of continuous symmetry on the classical background: application of
Bogoliubov group variables permits to take into account conservation laws
and avoid zero-mode problem. }

{\normalsize Conservation laws are the fundamental principles that connected
directly with symmetry properties of the system under consideration.
Bogoliubov method permits to take into account conservation laws precisely
and explicitly in the any order of perturbations theory construction. }

{\normalsize The main idea of the method consists in using new variables
that has sense of generators of system symmetry group. New variables are
cyclic while their canonical momenta turns out to be integrals of motion and
hence commute with Hamiltonian providing exact performance of conservation
laws. In the case of not weak interactions the perturbation theory as any
approximation method could violate exact performance of conservation laws
and we can't evaluate how our approximation of state vector is close to the
real state vector. Taking into account this circumstance we can state that
Bogoliubov method has significant advantage. }

{\normalsize The second dignity of this method application is the following:
we can avoid zero-mode problem that appears in the process of quantization
of the system that has continuous symmetry. }

{\normalsize In our resent articles we have developed the Bogoliubov method
for any relativistic system that allows existence of symplectic structures,
one of such systems is gravitational field. }

{\normalsize Now we would like to represent the scheme of quantization of
gravitational field on the classical background by means of Bogoliubov group
variables. }

\section{Space-time description}

{\normalsize We consider gravitational field in (3+1)-dimensioned formalism
that has been proposed by Arnowitt-Deser-Misner (ADM). }

{\normalsize Metrical tensor in this formalism looks like 

$$
g_{\alpha \beta }=
\left(
\matrix{
 -a^2+b^tb_t & {b^t} \cr
 {b_t}  & \gamma _{st}\cr
}
\right)
$$

here $\gamma _{st}$ is metrix of 3D-space in 4D-manyfold. }

{\normalsize Canonical momentum $\pi ^{st}$ is determined as usual: 
\[
\pi ^{st}=-\sqrt{\gamma }\left( K^{st}-\gamma ^{st}K\right) , 
\]
here 
\[
\sqrt{\gamma }=\sqrt{\det \left\| \gamma _{st}\right\| };\qquad
K_{tp}=-a\Gamma _{tp}^0; 
\]
}

{\normalsize 
\[
\Gamma _{tp}^s={\frac 12}\gamma ^{sp}\Gamma _{tlp}^{};\qquad \Gamma
_{tlp}^{}=\gamma _{pl,t}+\gamma _{pt,l}-\gamma _{tl,p}. 
\]
Denoting as usually 
\[
R_{\kappa \lambda }=\Gamma _{\kappa \lambda ,\sigma }^\sigma -\Gamma
_{\kappa \sigma ,\lambda }^\sigma +\Gamma _{\kappa \lambda }^\sigma \Gamma
_{\sigma \rho }^\rho -\Gamma _{\kappa \rho }^\sigma \Gamma _{\lambda \sigma
}^\rho , 
\]
we can represent the action of gravitational field 
\[
S=\int d^3x\sqrt{g}g^{\kappa \lambda }R_{\kappa \lambda } 
\]
in the following form: 
\[
S=\int d^3x\left( \pi ^{st}\gamma _{st,0}-aH-b_sH^s\right) , 
\]
here 
\[
H={\frac 1{\sqrt{\gamma }}}\left( \pi _{st}\pi ^{st}-{\frac 12}\pi ^2\right)
-\sqrt{\gamma }R, 
\]
\[
H^s=-2\pi _{;l}^{sl}. 
\]
}

{\normalsize Suppose that 4D manifold with given metrix permits to chose
space-like hypersurface }$\sum $ {\normalsize and to set normals field on
this hypersurface. Those normals are tangent to geodesic and determine time
coordinate. Hence geometry of 4D manifold could be described via Gaussian
coordinates:}

$$
g_{\alpha \beta }=
\left(
\matrix{
 -a^2 & 0 \cr
 0  & \gamma _{st}\cr
}
\right)
$$

We have to chose {\normalsize \ hypersurface }$\sum $ in according with the
foloowing principle:

Suppose that normal $\vec{n}$ is given in the {\normalsize \ hypersurface }$%
\sum $ at the

point $X.$ Let's chose arbitrary closed line $K\subset \sum .$ $\left(
X\subset K\right) $

Normal vector $\vec{n}$ have to coinside with it's original direction after
whole cycle motion along the line $K$- this is criterium of the hypersurface 
$\sum $ choice. {\normalsize Choquet-Bruhat showed that this criterium could
be realized via imposing of the Hamiltonian constraint:}

{\normalsize 
\begin{eqnarray*}
{\frac 1{\sqrt{\gamma }}}\left( \pi _{st}\pi ^{st}-{\frac 12}\pi ^2\right) 
\sqrt{\gamma }R &=&0 \\
&&
\end{eqnarray*}
}

and momentum constraint:

{\normalsize 
\[
\pi _{;l}^{sl}=0. 
\]
}

{\normalsize General principles of canonical formalism for the systems with
constraints leads us to the following statement\ (Lichnerovitch,
Choquet-Bruhat, Dirac, Antrowitt-Deser-Misner): }

If the evolutions equations

{\normalsize 
\[
\gamma _{st,0}={\frac{{2a}}{\sqrt{\gamma }}}\left( \pi _{st}-{\frac 12}%
\gamma _{st}\pi \right) +b_{s;t}+b_{t;s}; 
\]
\[
\pi _{,o}^{st}=-a\sqrt{\gamma }\left( R^{st}-{\frac 12}\gamma ^{st}R\right) +%
{\frac a{2\sqrt{\gamma }}}\left( \pi _{st}\pi ^{st}-{\frac 12}\pi ^2\right)
\gamma ^{st}+ 
\]
\[
-{\frac a{2\sqrt{\gamma }}}\left( \pi _l^s\pi ^{lt}-{\frac 12}\pi ^{st}\pi
\right) +\sqrt{\gamma }\left( \gamma ^{sl}c_{;l}^t-\gamma
^{st}c_{;l}^l\right) + 
\]
\[
+\left( \pi ^{st}b^l\right) _{;l}-\pi ^{sl}b_{;l}^t-\pi ^{lt}b_{;l}^s;\qquad
c^l=\gamma ^{ls}a_{;s} 
\]
are performed on the 3D-space, then in 4D manifold the Einstein equations
holds true: 
\[
R_{\mu \nu }-{\frac 12}g_{\mu \nu }R=0,\qquad \left(
R_{ikl;m}^n+R_{imk;l}^n+R_{ilm;k}^n=0\right) . 
\]
}

\section{Bogoliubov Group Variables}

{\normalsize Let's variables $x^{^{\prime }}$ are connected with $x$ by the
following group transformation: }

{\normalsize $x^{\prime }=f(x,a),$ $x^{\prime \prime }=f\left(
f(x,a),b\right) =f(x,c),$ $c=\varphi \left( a,b\right) .$}

Under variation of group parameters $a${\normalsize \ variation of
coordinates reads:}

{\normalsize $\left( \delta x^{\prime }\right) ^i=\xi _s^i(x^{\prime
})B_p^s(a)\delta a^p,$ }

{\normalsize here }

{\normalsize $i=0,1,2,3$- the number of coordinates, }

{\normalsize $p=1,...,r$ , where $r$ is a quantity of group gererators, }

{\normalsize $B_p^s(a)$ defines group properties. }

{\normalsize Note that conservation laws performance in curved space-time is
connected with Killing vectors existence, they are not straightforward
sequence of system space-time transformation invariance. }

{\normalsize In present case Bogoliubov transformation reconstruct
invariance with respect to transformation group, that has been violated due
to explicit extraction of classical field. }

{\normalsize It means the following: if we have made quantization in some
surface $\Sigma $ in definite moment, application of group variables permits
to state that we can move this surface $\Sigma $ with according to group
lows (including transformation that changes the time coordinate). }

Let's consider couples of functions {\normalsize $f_{st}(x),$ $f_n^{st}(x)$
and define Bogoliubov transformation as following: }

{\normalsize 
$$
f_{st}(x)=gv_{st}(x^{\prime })+u_{st}(x^{\prime }),\qquad
f_n^{st}(x)=gv_n^{st}(x^{\prime })+u_n^{st}(x^{\prime }),\eqno(2) 
$$
}

{\normalsize dimensionless parameter $g$ is assumed to be large, and group
parameters $a^p$ are independent new variables. }

{\normalsize The substitution $f_{st}(x)\rightarrow \{u_{st},(a)\}$ enlarge
the number of independent variables on $r$, so problem is how to formulate
invariant conditions, which we have to impose on functions $%
u_{st}(x^{^{\prime }})$ in order to equalize the number of independent
variables in the both part of equation (2). }

{\normalsize We consider systems in which there are invariant symplectic
forms that looks like the following:}

{\normalsize \ $\omega \left( u_{st},N^{stp}\right) =\int\limits_\Sigma
(u_n^{st}(x^{^{\prime }})N_{st}^p(x^{^{\prime }})-u_{st}(x^{^{\prime
}})N_n^{stp}(x^{^{\prime }}))d\sigma ,$}

{\normalsize here $\Sigma $ is some space-like surface. } Everywhere in the
article we mean summation with respect to all $s$ and $t.$

Additional conditions are:

{\normalsize 
$$
\omega \left( u_{st},N^{stp}\right) =0.\eqno(3)
$$
}

{\normalsize We chose some functions $N_{st}^p(x^{^{\prime }})$ ($p=1...r,$
the quantity functions is equal to the quantity of new independent
variables) }

{\normalsize It is possible to obtain equations, which define group
variables as functional of $f_{st}(x)$ and $f_n^{st}(x)$ on the $\Sigma $,
by substitution $u_{st}(x^{^{\prime }})$ in the additional conditions in
according with (2). }

{\normalsize Cause the forms $\omega (N_{st}^a,u_{st})$ are invariant with
respect to variations of $a^p$ one can obtain: }

{\normalsize \ 
\[
-\int\limits_\Sigma d\sigma \left( \delta f_{st}(x)N_n^{stk}(x^{\prime
})-\delta f_n^{st}(x)N_{st}^k(x^{\prime })\right) 
\]
$$
-gB_p^s(a)\delta a^p-B_p^r(a)\delta a^pR_r^s=0,\eqno(4)
$$
}

{\normalsize here\ }

{\normalsize $R_r^k=\int\limits_\Sigma d\sigma \xi _r^i(x^{\prime })\left(
N_{ni}^{stk}(x^{\prime })u_{st}(x^{\prime })-N_{sti}^k(x^{\prime
})u_n^{st}(x^{\prime })\right) .$ }

{\normalsize It is useful to formulate equation (4) in the differential
form: }

{\normalsize $\frac{\delta a^p}{\delta f_{st}(x)}=-\frac
1gA_k^p(a)N_n^{stk}(x^{\prime })-\frac 1g\frac{\delta a^q}{\delta f_{st}(x)}%
B_q^r(a)R_r^sA_s^p(a),$ }

{\normalsize $\frac{\delta a^p}{\delta f_n^{st}(x)}=\frac
1gA_k^p(a)N_{st}^k(x^{\prime })-\frac 1g\frac{\delta a^q}{\delta f_n^{st}(x)}%
B_q^r(a)R_r^sA_s^p(a),$ }

{\normalsize $A_s^p(a)$ denote the matrix inverse to $B_q^s(a):$ }

{\normalsize $B_q^s(a)A_s^p(a)=\delta _q^p.$}

Denote

$N_{st}^k(x^{\prime })T_k^l=D_{st}^l(x^{\prime })$

$N_n^{stk}(x^{\prime })T_k^l=D_n^{stl}(x^{\prime })$

{\normalsize where $T_s^l$ are the solution of the equation: }

{\normalsize $T_s^l=\delta _s^l-\frac 1gT_s^rR_r^l.$}

{\normalsize So we can state that }

{\normalsize $\frac{\delta a^p}{\delta f_{st}(x)}=\frac
1gA_l^p(a)D_n^{stl}(x^{\prime }),$ $\frac{\delta a^p}{\delta f_n^{st}(x)}%
=-\frac 1gA_l^p(a)D_{st}^l(x^{\prime }).$}

{\normalsize As a consequence of (3) we obtain linear dependence between
derivatives with respect to $u_{st}(x^{^{\prime }})$ and $%
u_n^{st}(x^{^{\prime }})$:}

{\normalsize \ $\int\limits_\Sigma d\sigma \left( M_{str}(x^{\prime })\frac
\delta {\delta u_{st}(x^{\prime })}+M_{nr}^{st}(x^{\prime })\frac \delta
{\delta u_n^{st}(x^{\prime })}\right) =0,$ }

{\normalsize where we define }

{\normalsize $\xi _r^i(x^{\prime })v_{sti}(x^{\prime })=M_{str}(x^{\prime
}), $ $\xi _r^i(x^{\prime })v_{ni}^{st}(x^{\prime })=M_{nr}^{st}(x^{\prime
}).$ }

(Here we demand the following relationship to be true:

$\omega \left( {M_{sta}}N_{st}^k\right) =\delta _a^k)$

{\normalsize Straightforward calculations give us $f_{st}(x)$ and $%
f_n^{st}(x)$ in the terms of new variables: }

{\normalsize $\frac \delta {\delta f_{st}(x)}=\frac \delta {\delta
u_{st}(x^{\prime })}+B_q^p(a)\frac{\delta a^q}{\delta f_{st}(x)}\left(
S_p+A_p^r(a)\frac \partial {\partial a^r}\right) ,$ }

{\normalsize $\frac \delta {\delta f_n^{st}(x)}=\frac \delta {\delta
u_n^{st}(x^{\prime })}+B_q^p(a)\frac{\delta a^q}{\delta f_n^{st}(x)}\left(
S_p+A_p^r(a)\frac \partial {\partial a^r}\right) ,$ }

{\normalsize here $S_p$ is defined as}

{\normalsize \ $-\int\limits_\Sigma d\sigma \xi _p^i(x^{\prime })\left(
u_{sti}(x^{\prime })\frac \delta {\delta u_{st}(x^{\prime
})}+u_{ni}^{st}(x^{\prime })\frac \delta {\delta u_n^{st}(x^{\prime
})}\right) =S_p$.}

\section{Secondary Quantization.}

{\normalsize The operators $\hat{q}_{st}(x)$ and $\hat{p}^{st}(x^{^{\prime
}})$:}

{\normalsize 
$$
\hat{q}_{st}(x)=\frac 1{\sqrt{2}}\left( f_{st}(x)+i\frac \delta {\delta
f_n^{st}(x)}\right) , \qquad \hat{p}^{st}(x)=\frac 1{\sqrt{2} }\left(
f_n^{st}(x)-i\frac \delta {\delta f_{st}(x)}\right) ,\eqno(5)
$$
}

{\normalsize are defined in the space $L$ of functionals $F$ where the
scalar product defines as 
\[
<F_1|F_2>=\int Df_{st}Df_n^{st}F_{1n}[f_{st},f_n^{st}]F_2[f_{st},f_n^{st}]. 
\]
The operators (5) are self-conjugated in this space. They are satisfy the
formal commutation relation: 
\[
\lbrack \hat{q}_{st}(x),\hat{p}^{st}(x^{^{\prime }})]=i\delta (x-x^{^{\prime
}}). 
\]
So we can treat $\hat{q}_{st}(x)$ and $\hat{p}^{st}(x)$ as operators of
coordinate and momentum of oscillators of field and we can develop the
secondary quantization scheme. But straightforward use of this procedure
leads us to the doubling of numbers of possible field states, because there
are self-conjugate operators }

{\normalsize $\tilde{q}_{st}(x)=\frac 1{\sqrt{2}}\left( f_{st}(x)-i\frac
\delta {\delta f_n^{st}(x)}\right) ,$ $\tilde{p}^{st}(x)=\frac 1{\sqrt{2}%
}\left( f_n^{st}(x)+i\frac \delta {\delta f_{st}(x)}\right) ,$ }

{\normalsize that are satisfy the same commutation relation and commute with 
$\hat{q}(x)$ and $\hat{p}(x)$. }

{\normalsize The operators $\tilde{p}(x)$ and $\tilde{q}$ also generate some
operators of coordinate and momentum, which are defined in the Fock space
orthogonal one described above. So quantization based on the (5) needs
further reduction of states number. One of the reduction method is the
holomorphic representation, for example. }

{\normalsize We can consider the space of functionals $F[z_{st},z^{st*}]$
isomorphic to the space $F[f_{st},f_n^{st}]$. If we define 
\[
z_{st}(x)=f_{st}(x)+if_n^{st}(x),\qquad z^{st*}(x)=f_{st}(x)-if_n^{st}(x), 
\]
so $\hat{q}_{st}(x)$ and $\hat{p}^{st}(x)$ can be defined as operators: }

{\normalsize $\hat{q}_{st}(x)=\frac 1{\sqrt{2}}\left( z_{st}(x)-\frac \delta
{\delta z^{st*}(x)}\right) ,$ $\hat{p}^{st}(x)=\frac 1{\sqrt{2}}\left(
z^{st*}(x)+\frac \delta {\delta z_{st}(x)}\right) .$}

{\normalsize In the space $F[z_{st},z^{st*}]$ reduction of states numbers is
made by the choice state vector in the form:}

{\normalsize $F=\left( exp\int\limits_\Sigma z_{st}(x)z^{st*}(x)d\sigma
\right) \Phi [z].$}

{\normalsize It is easy to see that vector }

{\normalsize $F_0=exp\left( \int\limits_\Sigma z_{st}(x)z^{st*}(x)d\sigma
\right) $}

{\normalsize is the vacuum of operators}

{\normalsize $\hat{q}_{st}(x)={\frac 1{\sqrt{2}}}\left( z_{st}(x)-{\frac
\delta {\delta z^{st*}(x)}}\right) ,$ $\hat{p}^{st}(x)={\frac 1{\sqrt{2}i}}%
\left( z^{st*}(x)+{\frac \delta {\delta z_{st}(x)}}\right) .$ So realization
of operators $\hat{q}_{st}(x)$ and $\hat{p}^{st}(x)$ is the holomorphic
representation. Due to conditions (3) that has appeared along with
Bogoliubov transformation, it is impossible to use holomorphic
representation for the reduction of states numbers straightway. }

{\normalsize And we use the following scheme: }

{\normalsize we use Bogoliubov 's transformation (2) and, in spite of
appearance of exceed states, we will develop scheme of perturbation theory.
Then reduction of states number will be made, so it will depend on dynamical
system equations. }

{\normalsize In the terms of new variables the operators of coordinate and
momentum reads: }

{\normalsize $\hat{q}_{st}(x)=\frac 1{\sqrt{2}}\left( gv_{st}(x^{\prime
})+u_{st}(x^{\prime })+i\frac \delta {\delta u_n^{st}(x^{\prime })}+B_q^p(a)%
\frac{\delta a^q}{\delta f_n^{st}(x)}\left( iS_p+iA_p^r(a)\frac \partial
{\partial a^r}\right) \right) ,$ }

{\normalsize $\hat{p}^{st}(x)=\frac 1{\sqrt{2}}\left( gv_n^{st}(x^{\prime
})+u_n^{st}(x^{\prime })-i\frac \delta {\delta u_{st}(x^{\prime })}-B_q^p(a)%
\frac{\delta a^q}{\delta f_{st}(x)}\left( iS_p+iA_p^r(a)\frac \partial
{\partial a^r}\right) \right) .$ }

{\normalsize Cause $g>>1$, operators ${\frac \partial {\partial \tau ^\alpha
}}$ enter in the $\hat{q}_{st}(x^{^{\prime }})$ and $\hat{p}%
^{st}(x^{^{\prime }})$ in the order $O\left( {\frac 1g}\right) $. With an
eye to increase the order of the velocity, it is necessary to make a
canonical transformation. Let's substitute state vector $\psi $ on the
vector:}

{\normalsize $\psi \longrightarrow e^{ig^2J}\psi ,$}

{\normalsize that accords to the substitution: }

{\normalsize 
$$
-iA_p^r(a)\frac \partial {\partial a^r}\longrightarrow g^2J_p-iA_p^r(a)\frac
\partial {\partial a^r}.\eqno(6)
$$
}

{\normalsize After canonical transformation the operators $\hat{p}^{st}(x)$
and $\hat{q}_{st}(x)$ become the following series:}

{\normalsize \ $\hat{q}_{st}=g\left( F_{st}(x^{\prime })+\frac 1g\hat{Q}%
_{st}(x^{\prime })+\frac 1{g^2}A_{st}(x^{\prime })\right) ,$}

{\normalsize \ $\hat{p}^{st}=g\left( F_n^{st}(x^{\prime })+\frac 1g\hat{P}%
^{st}(x^{\prime })+\frac 1{g^2}A_n^{st}(x^{\prime })\right) .$ }

Explicit expressions for the addends in the series are:

{\normalsize $F_{st}(x^{\prime })=\frac 1{\sqrt{2}}\left( v_{st}(x^{\prime
})+N_{st}^k(x^{\prime })J_k\right) ,$}

{\normalsize $F_n^{st}(x^{\prime })=\frac 1{\sqrt{2}}\left(
v_n^{st}(x^{\prime })+N_n^{stk}(x^{\prime })J_k\right) ,$ }

{\normalsize $\hat{Q}_{st}(x^{\prime })=\frac 1{\sqrt{2}}\left(
u_{st}(x^{\prime })+i\frac \delta {\delta u_n^{st}(x^{\prime
})}-N_{st}^k(x^{\prime })r_k\right) ,$ }

{\normalsize $\hat{P}^{st}(x^{\prime })=\frac 1{\sqrt{2}}\left(
u_n^{st}(x^{\prime })-i\frac \delta {\delta u_{st}(x^{\prime
})}-N_n^{stk}(x^{\prime })r_k\right) ,$ }

{\normalsize $A_{st}(x^{\prime })=\frac{\delta a^p}{\delta f_n^{st}(x)}%
\left( B_p^r(a)R_r^kr_k-iK_p\right) ,$ }

{\normalsize $A_n^{st}(x^{\prime })=\frac{\delta a^p}{\delta f_{st}(x)}%
\left( B_p^r(a)R_r^kr_k-iK_p\right) ,$ }

here

{\normalsize $T_c=K_c+R_c^ar_a,$ $K_p=B_p^q(a)S_q+\frac \partial {\partial
a^p},$ $r_k=R_k^pJ_p,$}

{\normalsize We define contravariant components of coordinate operator and
covariant component of momentum operator taking into account that they have
to satisfy the following relations: }

{\normalsize $\hat{q}^{sl}(x)\hat{q}_{st}(x)=\delta _t^l,$ $\hat{q}^{sl}(x)%
\hat{p}_{st}(x)=\hat{p}^{sl}(x)\hat{p}_{st}(x).$ }

{\normalsize It is possible if the operators reads:\ }

{\normalsize $\hat{q}^{st}(x)=g\left( F^{st}(x^{\prime })-\frac 1g\hat{Q}%
^{st}(x^{\prime })+\frac 1{g^2}B^{st}(x^{\prime })\right) ,$ }

{\normalsize $\hat{p}_{st}(x)=g\left( F_{nst}(x^{\prime })+\frac
1gS_{st}(x^{\prime })+\frac 1{g^2}D_{st}(x^{\prime })\right) ,$ }

{\normalsize and the addends are defined by the following way: }

{\normalsize $F^{sl}F_{n_{st}}=\delta _t^l,$ $\hat{Q}^{st}=F^{rt}\hat{Q}%
_{rl}F^{sl},$ }

{\normalsize $B^{st}=F^{sk}\hat{Q}_{kl}F^{lt}\hat{Q}%
_{sm}F^{mr}-F^{st}A_{sl}F^{lr},$ $F_n^{st}F_{sl}=F^{st}F_{n_{sl}},$}

{\normalsize $S_{kl}=F_n^{bt}\left( \hat{Q}_{bt}F_{kl}+\hat{Q}%
_{bl}F_{kt}\right) +F_{bl}\hat{P}^{st}F_{kt},$ }

{\normalsize $D_{pl}=A_{n_{pl}}+F_n^{st}\left(
F_{tp}A_{sl}+F_{sl}A_{tp}\right) +F_n^{sm}\hat{Q}_{sl}\hat{Q}_{mp}+2F_{tp}%
\hat{P}^{st}\hat{Q}_{sl}.$ }

{\normalsize Then action can be represented as series with respect to
inverted powers of coupling constant: }

{\normalsize 
$$
S=S_0+S_1+S_2.\eqno(7)
$$
}

\section{Perturbation Theory Construction{\protect\normalsize \ }}

{\normalsize Now we can quantize and substitute $u_{st}(x^{^{\prime }})$, $%
u_n^{st}(x^{^{\prime }})$ as following: }

{\normalsize $u_{st}(x^{^{\prime }})\mapsto \hat{q}^{st}(x)$, $%
u_n^{st}(x^{^{\prime }})\mapsto $\ $\hat{p}^{st}(x).$}

{\normalsize In the series (7) operators $S_0$ are $C$-numbers. }

{\normalsize Let's consider the higher order. }

{\normalsize $S_1=\int\limits_\Sigma A_{st}(x^{\prime })\hat{P}%
^{st}(x^{\prime })+B^{st}(x^{\prime })\hat{Q}_{st}(x^{\prime }),$ }

{\normalsize here explicit view of $A_{st}(x^{\prime })$ and $%
B^{st}(x^{\prime })$ are given in the Appendix 1, and operator $S_1$ is
linear with respect to $u_{st}(x^{^{\prime }})$, $u_n^{st}(x^{^{\prime }})$, 
${\frac \partial {\partial u_{st}(x^{^{\prime }})}}$, ${\frac \partial
{\partial u_n^{st}(x^{^{\prime }})}}$. There are no any normalizable
eigenvectors of these operators, so it is required to set them to zero for
perturbation theory construction. Let's explore if it is possible. }

{\normalsize We can write the action as :}

{\normalsize \ $S_1=\int\limits_\Sigma A_{st}\left( u_n^{st}-i\frac \delta
{\delta u_{st}}-N_n^{stk}r_k\right) +B^{st}\left( u_{st}+i\frac \delta
{\delta u_n^{st}}-N_{st}^kr_k\right) .$ }

We can represent actions as a sum:

{\normalsize $S_1=-i\int\limits_\Sigma A_{st}\frac \delta {\delta
u_{st}}-B^{st}\frac \delta {\delta u_n^{st}}+$}

{\normalsize $\int\limits_\Sigma A_{st}\left( u_n^{st}-N_n^{stk}r_k\right)
+B^{st}\left( u_{st}-N_{st}^kr_k\right) .$ }

Above we have got {\normalsize linear form with respect to ${\frac \delta
{\delta u_{st}(x^{^{\prime }})}}$ and ${\frac \delta {\delta
u_n^{st}(x^{^{\prime }})}}$ that is equal zero: }

{\normalsize $\int\limits_\Sigma d\sigma \left( M_{str}\frac \delta {\delta
u_{st}}+M_{nr}^{st}\frac \delta {\delta u_n^{st}}\right) =0.$ }

{\normalsize Linear form with respect to derivatives in $S_{-1}$ will be
equal zero if we demand $A_{st}(x^{^{\prime }})$ and $B^{st}(x^{^{\prime }})$
to be linearly connected with ${M_{sta}}(x^{^{\prime }})$ and $%
M_{na}^{st}(x^{^{\prime }})$: }

{\normalsize $A_{st}(x^{^{\prime }})=c^a{M_{sta}}(x^{^{\prime }}),$\ $%
B^{st}(x^{^{\prime }})=c^aM_{na}^{st}(x^{^{\prime }}).$\ }

{\normalsize Linear form with respect to $u_{st}(x^{^{\prime }})$ and $%
u_n^{st}(x^{^{\prime }})$ in the $S_1$ looks like}

{\normalsize \ $S_1=\int\limits_\Sigma A_{st}\left(
u_n^{st}-N_n^{stk}r_k\right) +B^{st}\left( u_{st}-N_{st}^kr_k\right) =$}

$c^a\int\limits_\Sigma \left( {M_{sta}}u_n^{st}-M_{na}^{st}u_{st}\right)
-r_k\int\limits_\Sigma \left( {M_{sta}}(x^{^{\prime
}})N_n^{stk}-M_{na}^{st}N_{st}^k\right) ${\normalsize $.$}

Taking into account that {\normalsize \ }

$\omega \left( {M_{sta}}N_{st}^k\right) =\delta _a^k$

and

$r_k=R_k^pJ_p=J_p${\normalsize $\int\limits_\Sigma d\sigma \xi
_r^i(x^{\prime })\left( N_{ni}^{stp}(x^{\prime })u_{st}(x^{\prime
})-N_{sti}^p(x^{\prime })u_n^{st}(x^{\prime })\right) .$ }

we state that the linear form with respect to {\normalsize $%
u_{st}(x^{^{\prime }})$ and $u_n^{st}(x^{^{\prime }})$ will be equal zero if
the following conditions are performed on $\Sigma $: $v_{st}(x^{\prime
})=J_kN_{st}^k(x^{\prime }),\qquad F_{st}(x^{^{\prime }})=\sqrt{2}%
v_{st}(x^{^{\prime }}).$}

Here we can obtain the expression for the parameter of canonical
transformation (6) (velocity of the classical part):

$J_k=\frac 1{\sqrt{2}}${\normalsize $\int\limits_\Sigma $}$F_n^{st}M{_{stk}-}%
F_{st}^{}M_{nk}^{st}.$

We've got that $c^a=\sqrt{2,}$ {\normalsize so we can state that linear form
of derivatives with respect to $u_{st}(x^{^{\prime }})$ and $%
u_n^{st}(x^{^{\prime }})$ in the operator $S_1$will be equal zero if the
following equations holds true on $\Sigma $: }

{\normalsize 
\[
{F_{stn}}=\frac{2a}{\sqrt{F}}\left( F_{stn}-\frac 12F_nF_{st}\right) ,
\]
}

{\normalsize 
$$
F_{nn}^{st}=\frac a{2\sqrt{F}}\left( F_{lkn}F_n^{lk}-\frac 12F_n^2\right)
F^{st}-\frac{2a}{\sqrt{F}}\left( F_n^{st}F_n^{kl}F_{stn}-\frac
12F_nF_n^{st}\right) - \eqno(8)
$$
}

{\normalsize 
\[
-a\sqrt{F}\left( R^{st}-\frac 12F_{}^{st}R\right) -\sqrt{F}\left(
F^{sl}c_{;l}^t-F^{st}c_{;l}^l\right) ,
\]
}

These equations could be treated as evolution equations.

{\normalsize Herandafter we assume $F_{st}(x)$ to be solution of the
equations (8), and $F_{st}(x^{^{\prime }})$ and $F_n^{st}(x^{^{\prime }})$
on $\Sigma $ are the solution of the Cauchy problem on $\Sigma $, so we can
state that on the 3D manifold the evolution equation holds true. The
constraint equations }

$\frac 1{\sqrt{F}}\left( F_n^{st}F_{stn}-\frac 12F_n^2\right) -\sqrt{F}%
R(F)=0,$ $F_{;l}^{sl}=0$

{\normalsize we obtain as a conditions on the choice of $\Sigma $ surface.
(See (1))}

{\normalsize So we can state that Einstein equations performance is
necessary condition for the perturbation theory to be applicable. We would
like to underline that Einstein equations has been obtained in the process
of perturbation theory construction as a condition of validity, not as a
sequence of variational principle.}

\section{Conclusion}

We applied Bogoliubov transformation to the quantization of gravitational
field in the neighbourhood of nontrivial classical component, that permitted
us to avoid zero-mode problem.

Einstein equations for the classical component has been obtained as a
necessary condition for the perturbation theory to be applicable.

The expression for quantum corrections of the field operator and explicit
view of state si the task of the next article.

\section{\protect\normalsize Appendix 1 }

{\normalsize Let's consider expansion of Hamiltonian into the series with
respect to inverted powers of coupling constant: }

{\normalsize $H=\frac 1{\sqrt{\gamma }}\left( \pi _{st}\pi ^{st}-\frac 12\pi
^2\right) -\sqrt{\gamma }R$ }

{\normalsize We have the order that depends only from classical part of the
field:\ }

{\normalsize $H_0=\frac 1{\sqrt{F}}\left( F_{n_{st}}F_n^{st}-\frac
12F_n^2\right) -\sqrt{F}R(F),$ }

{\normalsize the following order is linear with respect to quantum addend
and derivatives with respect to quantum addends: }

{\normalsize $H_1=\frac 1{\sqrt{F}}\left( 
\begin{array}{c}
\frac 12\left( F_{n_{kl}}F_n^{kl}-\frac 12F_n^2\right) F^{st}\hat{Q}_{st}+
\\ 
\left( \hat{P}^{st}F_{n_{st}}+F_n^{st}S_{st}\right) -F_n\left( \hat{P}%
^{st}F_{n_{st}}+\hat{Q}_{st}F_n^{st}\right)
\end{array}
\right) $ }

{\normalsize $-\sqrt{F}\left( \frac 12F_{}^{st}R_{st}(F)\hat{Q}_{st}-\hat{Q}%
_{}^{st}R_{st}(F)+F_{}^{st}R_{st}(F,\hat{Q})\right) ,$ }

{\normalsize and the next order depends on 2dn order of quantum addend and
contains Bogoliubov variables: }

{\normalsize $H_2=\frac 1{\sqrt{F}}\left( 
\begin{array}{c}
-\frac 12\left( F_{n_{kl}}F_n^{kl}-\frac 12F_n^2\right) F^{st}A_{st}+ \\ 
\left( A_n^{st}F_{n_{st}}+F_n^{st}D_{st}+\hat{P}^{st}S_{st}\right) - \\ 
F_n\left( A_n+A_{st}F_n^{st}+\hat{Q}_{st}\hat{P}^{st}\right) -\frac 12F_n(%
\hat{P}+F_n^{st}\hat{Q}_{st})^2 \\ 
\frac 12F^{st}\hat{Q}_{st}\left( \left( \hat{P}%
^{st}F_{n_{st}}+S_{st}F_n^{st}\right) -F_n(\hat{P}+F_n^{st}\hat{Q}%
_{st})\right)
\end{array}
\right) -$ }

{\normalsize $-\sqrt{F}\left( 
\begin{array}{c}
\frac 12RF_{}^{st}A_{st}+R(F,\hat{Q},A)+B_{}^{st}R_{st}-\hat{Q}%
_{}^{st}R_{st}(F,\hat{Q}) \\ 
+\frac 12\left( F_{}^{st}R(F,\hat{Q})-\hat{Q}_{}^{st}R_{st}\right) F_{}^{st}%
\hat{Q}_{st}
\end{array}
\right) .$ }

{\normalsize Lets consider the addend $a\sqrt{F}F_{}^{st}R_{st}(F,\hat{Q})$ .%
}

{\normalsize Recall that:\ $\Gamma _{lt}^s=\gamma ^{sp}\Gamma _{ltp},$ }

{\normalsize and Kristoffel symbols are the series: \ }

{\normalsize $\Gamma _{ltp}=\Gamma _{ltp}(F)+\frac 1g\Gamma _{ltp}(\hat{Q}%
)+\frac 1{g^2}\Gamma _{ltp}(A),$ }

{\normalsize so $\Gamma _{lt}^s=\Gamma _{lt}^s(F)+\frac 1g\Gamma _{lt}^s(%
\hat{Q}),$ }

{\normalsize where $\Gamma _{lt}^s(\hat{Q})=\left( F^{sp}\Gamma _{ltp}(\hat{Q%
})-\hat{Q}^{sp}\Gamma _{ltp}(F)\right) .$ }

{\normalsize Taking into account that }

{\normalsize $\hat{Q}_{tp_{;l}}=\hat{Q}_{tp_{,l}}-\Gamma _{lt}^m(F)\hat{Q}%
_{mp}-\Gamma _{lp}^m(F)\hat{Q}_{mt},$ }

{\normalsize we note: }

{\normalsize $\Gamma _{ltp}(\hat{Q})=\frac 12\left( \hat{Q}_{tp;_l}+\hat{Q}%
_{lp;_t}-\hat{Q}_{lt;_p}\right) -2\hat{Q}_{mp}\Gamma _{lt}^m(F),$ }

{\normalsize so we can state that }

{\normalsize $\Gamma _{lt}^s(\hat{Q})=\frac 12F^{sp}\left( \hat{Q}_{tp;_l}+%
\hat{Q}_{lp;_t}-\hat{Q}_{lt;_p}\right) ,$ }

{\normalsize and hence Ricci tensor is the series too: }

{\normalsize $R_{st}=R_{st}(F)+\frac 1gR_{st}(F,\hat{Q})+\frac
1{g^2}R_{st}(F,\hat{Q},A),$ }

{\normalsize where }

{\normalsize $R_{st}(F,\hat{Q})=\Gamma _{st;_l}^l(\hat{Q})-\Gamma _{st;_t}^l(%
\hat{Q})+$ }

{\normalsize $+\Gamma _{st}^l(\hat{Q})\Gamma _{lm}^m(F)+\Gamma
_{st}^l(F)\Gamma _{lm}^m(\hat{Q})-\Gamma _{sl}^m(\hat{Q})\Gamma
_{mt}^l(F)-\Gamma _{sl}^m(F)\Gamma _{mt}^l(F),$ }

{\normalsize so that }

{\normalsize $R_{st}(F,\hat{Q})=\left( \Gamma _{st}^l(\hat{Q})\right)
_{;_l}-\left( \Gamma _{sl}^l(\hat{Q})\right) _{;t}$ }

{\normalsize and addend under consideration is }

{\normalsize $a\sqrt{F}F_{}^{st}R_{st}(F,\hat{Q})=$ }

{\normalsize $\sqrt{F}\left( aF_{}^{st}\Gamma _{st}^l(\hat{Q})\right) ,_l-%
\sqrt{F}\left( aF_{}^{st}\Gamma _{sl}^l(\hat{Q})\right) ,_t+$ }

{\normalsize $+\sqrt{F}F_{}^{st}a_{;t}\Gamma _{sl}^l(\hat{Q})-\sqrt{F}%
F_{}^{st}a_{;t}\Gamma _{sl}^l(\hat{Q}),$ }

{\normalsize here we note }

{\normalsize $c^s=a_{;t}F^{st}.$}

{\normalsize Let's consider the form: }

{\normalsize $\sqrt{F}F_{}^{st}a_{;t}\Gamma _{sl}^l(\hat{Q})=\sqrt{F}%
c_{}^s\frac 12F^{lp}\left( \hat{Q}_{sp;_l}+\hat{Q}_{lp;_s}-\hat{Q}%
_{sl;_p}\right) .$ }

{\normalsize Taking into account that the expression is equal zero }

{\normalsize $F^{lp}\left( \hat{Q}_{sp;_l}-\hat{Q}_{sl;_p}\right) =0$ }

{\normalsize because it contains product of symmetric and antisymmetric
tensors. Analogously }

{\normalsize $\sqrt{F}F_{}^{st}a_{;t}\Gamma _{sl}^l(\hat{Q})=\frac 12\left( 
\sqrt{F}c_{}^sF^{lp}\hat{Q}_{lp}\right) _{;s}-\frac 12\sqrt{F}F^{lp}\hat{Q}%
_{lp}c_{;s}^s$ }

{\normalsize and }

{\normalsize $\sqrt{F}F_{}^{st}a_{;l}\Gamma _{st}^l(\hat{Q})=$ $Div+\frac 12%
\sqrt{F}F^{lp}\hat{Q}_{lp}c_{;s}^s-\sqrt{F}F^{sl}\hat{Q}_{lp}c_{;s}^p$ ,}

{\normalsize and finally the addend reads: }

{\normalsize $a\sqrt{F}F_{}^{st}R_{st}(F,\hat{Q})=Div+\sqrt{F}\hat{Q}%
_{st}\left( F^{sp}c_{;p}^t-F^{st}c_{;p}^p\right) .$ }

{\normalsize Analogously we can state: }

{\normalsize $a\sqrt{F}F_{}^{st}R_{st}(F,\hat{Q},A)=Div+\sqrt{F}A_{st}\left(
F^{sp}c_{;p}^t-F^{st}c_{;p}^p\right) $ }

{\normalsize Lets' consider the following order of the addend: }

{\normalsize $a\sqrt{F}F_{}^{st}F_{}^{st}\hat{Q}_{st}R_{st}(F,\hat{Q})=$ }

{\normalsize $Div+\sqrt{F}F^{st}\left( a\hat{Q}\right) _{;t}\Gamma _{sl}^l(%
\hat{Q})-\sqrt{F}F^{st}\left( a\hat{Q}\right) _{;l}\Gamma _{st}^l(\hat{Q}).$ 
}

{\normalsize Denote }

{\normalsize $r^s=\left( a\hat{Q}\right) _{;t}F^{st},$ }

{\normalsize so straightforward calculations shows us that }

{\normalsize $a\sqrt{F}F_{}^{st}\hat{Q}R_{st}(F,\hat{Q})=\sqrt{F}\hat{Q}%
\left( F^{sp}r_{;p}^t-F^{st}r_{;p}^p\right) =$ }

{\normalsize $\sqrt{F}\left( F^{sp}c_{;p}^t-F^{st}c_{;p}^p\right) \hat{Q}%
_{st}\hat{Q}_{st}F^{st}+a\sqrt{F}F^{st}\left(
F^{sp}F^{tr}-F^{st}F^{pr}\right) \hat{Q}_{st,_{pr}},$}

{\normalsize and we can state that the addend is looks like the following: }

{\normalsize $a\sqrt{F}\hat{Q}_{}^{st}R_{st}(F,\hat{Q})=\sqrt{F}\left( a\hat{%
Q}^{st}\right) _{;t}\Gamma _{sl}^l(\hat{Q})-\sqrt{F}\left( a\hat{Q}%
^{st}\right) _{;l}\Gamma _{st}^l(\hat{Q})=$ }

{\normalsize $\sqrt{F}\left( F^{sp}a_{;tp}\hat{Q}^{st}-F^{st}a_{;tp}\hat{Q}%
^{pt}\right) \hat{Q}_{st}+a\sqrt{F}F^{st}\left( F^{sp}\hat{Q}%
_{;tp}^{st}-F^{st}\hat{Q}_{;tp}^{pt}\right) \hat{Q}_{st}.$}

So we can state that

{\normalsize $S_1=\hat{P}^{st}(x^{\prime })F_{n_{st}}(x^{\prime
})+F_n^{st}(x^{\prime })\hat{Q}_{n_{st}}(x^{\prime })+aH_1(F,\hat{Q})=$ }

{\normalsize $=\int\limits_\Sigma A_{st}(x^{\prime })\hat{P}^{st}(x^{\prime
})+B^{st}(x^{\prime })\hat{Q}_{st}(x^{\prime }),$ }

{\normalsize $A_{st}=\frac{2a}{\sqrt{F}}\left( F_{n_{st}}-\frac
12F_nF_{st}\right) ,$ }

{\normalsize $B^{st}=\frac a{2\sqrt{F}}\left( F_{n_{kl}}F_n^{kl}-\frac
12F_n^2\right) F^{st}-\frac{2a}{\sqrt{F}}\left(
F_n^{st}F_n^{kl}F_{n_{st}}-\frac 12F_nF_n^{st}\right) -$ }

{\normalsize $-a\sqrt{F}\left( R^{st}-\frac 12F_{}^{st}R\right) -\sqrt{F}%
\left( F^{sl}c_{;l}^t-F^{st}c_{;l}^l\right) ,$ }

\section{ References:}

\subsection{General principles:}

\textrm{$[I]$ N.N.Bogoliubov, D.V.Shirkov INTRODUCTION to the THEORY 
of QUANTIZED FIELDS, (New-York - London, Interscience Publishes 1958)}

\textrm{$[II]$ C.W.Misner, K.S.Thorne, J.A.Wheeler GRAVITAION. (W.H.Freeman
and Company, San Francisco, 1973)}

\textrm{$[III]$ N.D.Birrell, P.C.W.Davies QUANTUM FIELD in the  
 CURVED SPACE (Cambrige University Press, 1982) }

\textrm{$[IV]$ R.Rajaraman. SOLITONS and INSTANTONS: an Introduction to 
Solitons and Instantins in Quantum Field Theory (North-Holland, 1982)}

\textrm{\ }

\subsection{Bogoliubov Transformation:}

\textrm{$[1]$ N.N.Bogoliubov // the Ukrainian Mathematical Journal. 1950. '.
2. '. 3-24. }

\textrm{$[2]$ E.P.Solodovnikova, A.N.Tavkhelidze, O.A.Khrustalev //
Teor.Mat.Fiz. 1972. T.10. }

\textrm{. 162-181; T. 11. p. 317-330; 1973. T. 12. p. 164-178. }

\textrm{$[3]$ O.D.Timofeevskaya // Teor.Mat.Fiz 1983. T. 54. p. 464-468. }

\textrm{$[4]$ N.H.Christ, T.D.Lee // Phys. Rev. D. 1975. V. 12. P.
1606-1627. }

\textrm{$[5]$ E.Tomboulis // Phys. Rev. D. 1975. V. 12. P. 1678-1683. }

\textrm{$[6]$ M.Greutz // Phys. Rev. D. 1975. V. 12. P. 3126-3144. }

\textrm{$[7]$ K.A.Sveshnikov // Teor.Mat.Fiz. 1985. T. 55. p. 361-384; 1988.
T. 74. '. }

\textrm{$[8]$ O.A.Khrustalev, M.V.Tchitchikina // Teor.Mat.Fiz. 1997. T.
111. N2.p. 242-251; }

\subsection{Gravity and Quantum Gravity:}

\textrm{$[9]$ R.Arnowitt, S.Deser, C.W.Misner//Phys. Rev 1960 V.120 N.1
p.863-870}

\textrm{$[10]$ E.T.Newman, L.Tamburino, T.Unti//J. Math. Phys. 1963 V.7 N.5
p 915-923}

\textrm{$[11]$ E.T.Newman, L.Penrose//J. Math. Phys. 1966 V.7 N.5 p.863-870}

\textrm{$[12]$ W.Misner//J. Math. Phys. 1967 V.4 N.7 p.924-937}

\textrm{$[13]$ Y.Choquet-Bruhat, R.Georch//Commun. Math. Phys. 1969 N.14
p.329-335}

\textrm{$[14]$ J.W.York//Phys. Rev. Letters 1971 V.26 N.26 p.1656-1658}

\textrm{$[15]$ J.W.York//J. Math. Phys. 1972 V.13 N.2 p.125-130}

\textrm{$[16]$ A.E.Fisher, J.E.Marsden // J. Math. Phys. 1972. V.13 N.4 p.
546-568 }

\textrm{$[17]$ K.Kuchar//J. Math. Phys. 1972 V.13 N.5 p.768-781}

\textrm{$[18]$ S.A.Fulling // Phys. Rev D 1973. V.7 N.10 p. 2850-2862}

\textrm{$[19]$ N.O.Murchadha, J.W.Jork // Phys. Rev D 1974. V.10 N.2 p.
428-446}

\textrm{$[20]$ T.Regge, C.Teiteloim // Annals of Phys. 1974. N.88 p.286-318 }

\textrm{$[21]$ D.Cramer // ACTA\ PHYSICA POLONICA 1975. N.4 p.467-478}

\textrm{$[22]$ P.Cordero, C.Teiteloim //Annals of Phys. 1976. N.100 p.607-631%
}

\textrm{$[23]$ A.E.Fisher, J.E.Marsden//Gen. Relat. and Grav. 1976 V.7 N.12
p.915-920}

\textrm{$[24]$ K.Kuchar//J. Math. Phys. 1976 V.17 N.5 p.777-780}

\textrm{$[25]$ G.W.Gibbsons, D.N.Page, C.N.Pope // Commun. Math.Phys. 1990.
N127 p. 529-553 }

\textrm{$[26]$ R.Gomes, P.Laguna, Ph.Papadopoulos, J.Winicour//
gr-qc/9603060 1996 }

\textrm{$[27]$ C.Rovelli// gr-qc/0006061 2000 V.2 }

\end{document}